\documentclass[fleqn,usenatbib]{mnras}

\usepackage{newtxtext,newtxmath}

\usepackage[T1]{fontenc}

\DeclareRobustCommand{\VAN}[3]{#2}
\let\VANthebibliography\thebibliography
\def\thebibliography{\DeclareRobustCommand{\VAN}[3]{##3}\VANthebibliography}

\usepackage{graphicx}	% Including figure files
\usepackage{amsmath}	% Advanced maths commands
\usepackage{supertabular}
\defcitealias{Weltevrede2006}{W06}
\defcitealias{Malofeev2018}{M18}

\title[11 new RRAT in PUMPS]{Search for rotating radio transients in PUMPS}

\author[S.A. Tyul'bashev et al.]{
S. A. Tyul'bashev,$^{1}$\thanks{E-mail: serg@prao.ru (SAT)}
M. A. Kitaeva,$^{1}$
E.A. Brylyakova,$^{1}$
V.S. Tyul'bashev,$^{2}$
G.E. Tyul'basheva,$^{3}$
\\
$^{1}$ P.N. Lebedev Physical Institute of the Russian Academy of Sciences, Astro Space Center, Pushchino Radio Astronomy Observatory,\\
Radiotelescopnaya 1a, Moscow reg., Pushchino 142290, Russia \\
$^{2}$ Yandex LLC 16, Leo Tolstoy St., Moscow 119021, Russia \\
$^{3}$ Institute of Mathematical Problems of Biology, brunch of Keldysh Institute of Applied Mathematics,\\  
Vitkevich 1, Moscow reg., Pushchino 142290, Russia\\
}

\date{September 14, 2023}

\pubyear{October, 2023}

\begin{document}
%\label{firstpage}
\maketitle

\begin{abstract}
A search for pulsed radiation at a frequency of 111 MHz in the direction of 116 RRAT candidates was carried out. For the search, archival data obtained on a meridian 128-beam radio telescope, a Large Phased Array (LPA), was used. For each candidate, about six days of observations were accumulated over an interval of eight years. Eleven new RRATs have been discovered. It was possible to estimate periods for six of them, and to construct average profiles for four of them. Some of the candidates turned out to be known pulsars observed in the side lobes of the radio telescope and interference. For the part of the candidates could not find pulses with a signal-to-noise ratio of more than seven, and their nature remains unknown.
\end{abstract}

%: J0034-0721, J0323+3944, J0528+2200, J0611+3016, J0826+2637, J1239+2453, J1921+2153, J2234+2114


%%%%%%%%%%%%%%%%%%%%%%%%%%%%%%%%%%%%%%%%%%%%%%%%%%

%%%%%%%%%%%%%%%%% BODY OF PAPER %%%%%%%%%%%%%%%%%%

\section{Introduction}

Rotating Radio Transients (RRAT) — this is a special kind of pulsars, discovered by \citeauthor{McLaughlin2006} (\citeyear{McLaughlin2006})  in the search for dispersed pulses (pulses recorded first at high and then at low frequencies). The peculiarity of RRAT is that, unlike classical second pulsars, the time of occurrence of the next pulse is unpredictable, and it can take from tens of seconds to tens of hours between successive pulses (\citeauthor{McLaughlin2006}, \citeyear{McLaughlin2006}; \citeauthor{Logvinenko2020}, \citeyear{Logvinenko2020}). 
To search for RRAT, as a rule, direct pulse search methods are used with a search for possible dispersion measure (DM) and possible pulse widths.

According to \citeauthor{Keane2008} (\citeyear{Keane2008}), the amount of RRAT in the galaxy should be twice as large as the number of ordinary second pulsars. In catalogs containing RRAT (https://www.atnf.csiro.au/people/pulsar/psrcat/; http://astro.phys.wvu.edu/rratalog/; https://bsa-analytics.prao.ru/ en/transients/rrat/ ), as well as in recent works, there are approximately 250 RRATs. Of the recent works, we note the preprint 2023, which published 76 RRATs found on the 500-m FAST telescope (\citeauthor{Zhou2023}, \citeyear{Zhou2023}). The ATNF pulsar catalog (\citeauthor{Manchester2005}, \citeyear{Manchester2005}) contains almost  3,400 pulsars, which means that the proportion of transients is less than 10\% of the total number of pulsars.  Obviously, there is a large shortage of RRAT.

There is no clear understanding of the nature of RRAT yet. There are a number of phenomenological hypotheses explaining the observed properties of RRAT (\citeauthor{Weltevrede2006}, \citeyear{Weltevrede2006}; \citeauthor{Zhang2007}, \citeyear{Zhang2007}; \citeauthor{Wang2007}, \citeyear{Wang2007}; \citeauthor{Brylyakova2021}, \citeyear{Brylyakova2021}; \citeauthor{Tyulbashev2021}, \citeyear{Tyulbashev2021}). During the RRAT studies, it turned out that some of them are pulsars, in which the pulse energy distribution is lognormal with large deviations of the pulse energy from the average value. Thus, their energy distribution of impulses has a long ''tail``. For such transients, observers see the brightest pulses from the ''tail`` of the distribution, but do not detect regular (periodic) radiation due to insufficient sensitivity during the observation session.

Some of the transients appear to be pulsars with giant pulses. Pulsars with giant pulses are characterized by a power-law distribution of pulses at the tail of the distribution. It is the power distribution of pulses that is registered in the RRAT series. This indicates the likely detection of giant pulses. For such RRATs, observers can
register giant pulses, but are not able to see regular radiation. Obviously, for pulsars with a long tail of the energy distribution of pulses and pulsars with giant pulses, an average profile should accumulate during long observation sessions. Some of the transients are pulsars with a high proportion of nullings, that is, with a large number of missed pulses. Pulsars with nullings between individual pulses do not have regular radiation, and therefore pulsars with a high proportion of nullings are not detected by standard periodic signal search methods. For
pulsars with a high proportion of nullings, an average profile cannot be accumulated during long observation sessions.

There are also intermittent pulsars. They have periods of activity, followed by periods of absence of pulses. If
the activity periods are small, then such pulsars may be missed during a standard search using power spectra or periodograms.

Special studies have been conducted only for a couple dozen bright RRATs with hundreds and thousands of pulses found (see, for example, \citeauthor{Reynolds2006} (\citeyear{Reynolds2006}); \citeauthor{Karastergiou2009} (\citeyear{Karastergiou2009}); \citeauthor{Keane2011} (\citeyear{Keane2011}); \citeauthor{Palliyaguru2011} (\citeyear{Palliyaguru2011}); \citeauthor{Bhattacharyya2018} (\citeyear{Bhattacharyya2018}); \citeauthor{Brylyakova2021} (\citeyear{Brylyakova2021}); \citeauthor{Smirnova2022} (\citeyear{Smirnova2022}); \citeauthor{Chen2022} (\citeyear{Chen2022}); \citeauthor{Hsu2023} (\citeyear{Hsu2023}); \citeauthor{Zhang2023} (\citeyear{Zhang2023})). Due to the small number of investigated RRATs, a selection effects are not excluded, since for weak RRATs it is difficult to determine their nature and, accordingly, their share in the total number of transients.

The search of RRAT is difficult for two reasons. Firstly, the unpredictability of the appearance of pulses in time leads to the need to allocate a lot of observational time for their study. Secondly, the RRAT pulses are weak in general, therefore, in addition to the observation time, telescopes providing the highest sensitivity are also needed. So, in the study of transients J1538+2345, J1854+0306, J1913+1330 on telescopes with a diameter of 64, 76, 100, 300 m (Parkes, Jodrell Bank, Green Bank, Aresibo) and on the LOFAR distributed interferometric system, from units to tens of pulses per hour were recorded. At the same time, observations on the 500-m FAST telescope (Guaizhou) of the same
transients show the rate of arrival of pulses from 1.5 to 100 times higher (\citeauthor{Lu2019}, \citeyear{Lu2019}).

In total, with using Large phased array (LPA) telescope at the Pushchino Radio Astronomy Observatory (PRAO) in a number of searches, 48 RRATs were found (https://bsa-analytics.prao.ru/en/transients/rrat/). At the same time, during the past searches, some of the candidates were rejected by us, since the signal-to-noise ratio (S/N) in their profile turned out to be less than six. To visually search for pulses, we built dynamic spectra, i.e. drawings, on which frequency is represented on a vertical scale, and time is represented on a horizontal scale. The flux density on the spectra is displayed in individual pixels of the drawing in shades of color from white to black. A bright pulse on the dynamic spectrum is visible as a diagonal strip, and the slope of this strip is determined by the quadratic  dependence of the pulse arrival time on the frequency and indicates the dispersion measure (DM). After compensation of the DM and addition of all frequency channels, a pulse profile can be obtained. For weak pulses, no diagonal stripe is visible in the dynamic spectrum, whereas the pulse profile is still clearly distinguishable. In the current work, we are talking about the verification carried out for previously found weak pulses in archival data accumulated over eight years.

\begin{figure}
	\includegraphics[width=\columnwidth]{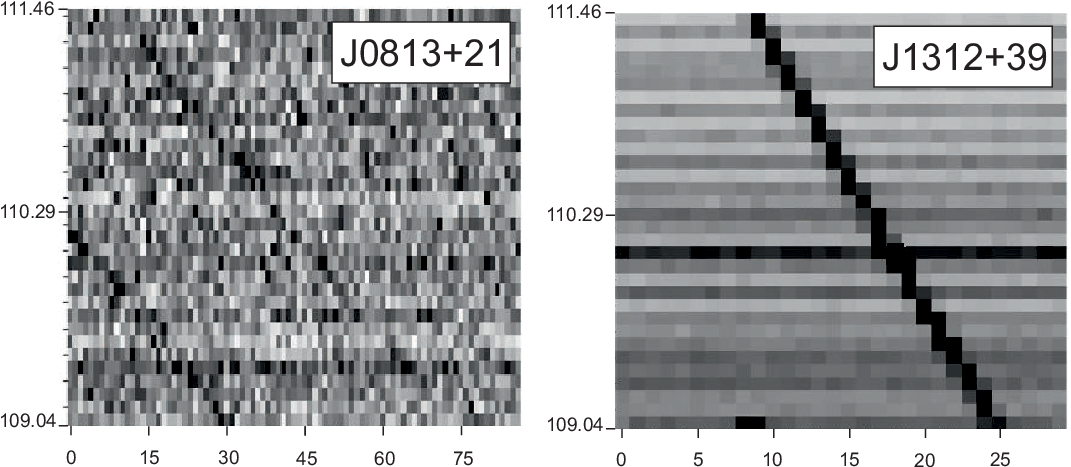}
	\caption{Dynamic spectra of two new transients. Time is on the horizontal axis. One point corresponds to 12.5 ms. The vertical axis is the frequency in MHz.}
    \label{fig:fig1}
\end{figure}

\section{OBSERVATIONS AND PROCESSING}

In 2012, the modernization of the LPA radio telescope was completed (\citeauthor{Shishov2016}, \citeyear{Shishov2016}), and it became possible to implement four independent radio telescopes based on a single antenna field. Now there are two radio telescopes operating in the PRAO, one of which (LPA3) has been conducting round-the-clock daily observations (since August 2014) in 96 spatial rays located along the meridian plane. Since 2022, 128 beams have been observed simultaneously on LPA3, covering an area of $\approx$ 50 square degrees in the sky. Monitoring observations on LPA3 are carried out at a frequency of 111 MHz, in the range of 2.5 MHz, divided into 32 frequency channels with a width of 78 kHz. The duration of the source recording is determined by the size of the LPA directional pattern and depends on the declination. Since the LPA is a meridian instrument, any source in the sky can be observed once a day. The duration of recording in the direction of the zenith is approximately 3.5 minutes per session, which corresponds to the accumulation time at each point in the sky of about six days over eight years of observations. The sampling time of the point is 12.5 ms. The data obtained in the 32-channel mode is used to search for pulsars and transients (\citeauthor{Tyulbashev2016}, \citeyear{Tyulbashev2016}; \citeauthor{Tyulbashev2018a}, \citeyear{Tyulbashev2018a}).

According to the work of \citeauthor{Tyulbashev2018a} (\citeyear{Tyulbashev2018a}), the sensitivity of LPA3 in the search for pulsed radio emission is approximately equal to 2.1 Jy for S/N, equal to seven, if the pulse duration coincides with the time interval of digitization of the signal. Since the observed RRAT pulse widths can be 40-50 ms (\citeauthor{Tyulbashev2018a}, \citeyear{Tyulbashev2018a}), for a part RRAT with wide pulses sensitivity can be improved to one Jy if you average the initial data before searching for pulses.

The volume of monitoring data accumulated since 2014 is approaching 250 terabytes, so the main problems in  processing observations are fast access to data and processing speed. In 2022, two servers with disk shelves of 24 disks appeared at the observatory. Raid arrays are implemented on the shelves. Each of the servers currently has two terabytes of RAM. A part of the operational memory is reserved for virtual disks, on which intermediate counting results are recorded. The final results of data processing are recorded on hard drives, which occupy small amounts of memory. The servers made it possible to organize both fast access to data and the fastest possible processing.

116 candidates with declination of $+21^o < \delta < +42^o$ were selected for verification. All candidates show signs of RRATs: the height of the pulse profile is 5-6, each candidate is observed in one beam, no obvious interference is visible in the recordings, signs of dispersed pulses are observed on the dynamic spectra, i.e. black pixels are visible along the diagonal line. For all candidates, there were estimates of coordinates, pulse width, and dispersion measures from previous searches. 

When searching for a dispersed signal of the candidate's RRAT, a number of standard procedures were performed: the baseline was subtracted in each frequency channel; dispersion measure were taken near the expected one with subsequent addition of frequency channels; the rms deviations of noise ($\sigma_{n}$) in the direction of the candidate were determined; the signal amplitude (A) for each DM being tested; signals having a S/N > 6 (S/N = $A/\sigma_{n}$) were recorded; different pulse widths were searched, assuming that the width of the pulse profile could be 1, 2, 4, 8, 16 points (12.5–200 ms). All the standard procedures described above were carried out for each iterated width. The dynamic spectra and pulse profiles were visually monitored for all the pulses found. In addition to interference monitoring, visual viewing allows you to detect additional pulses having a S/N < 6. These pulses were used to clarify the period
of pulsars.

\begin{figure*}
	\includegraphics[width=\textwidth]{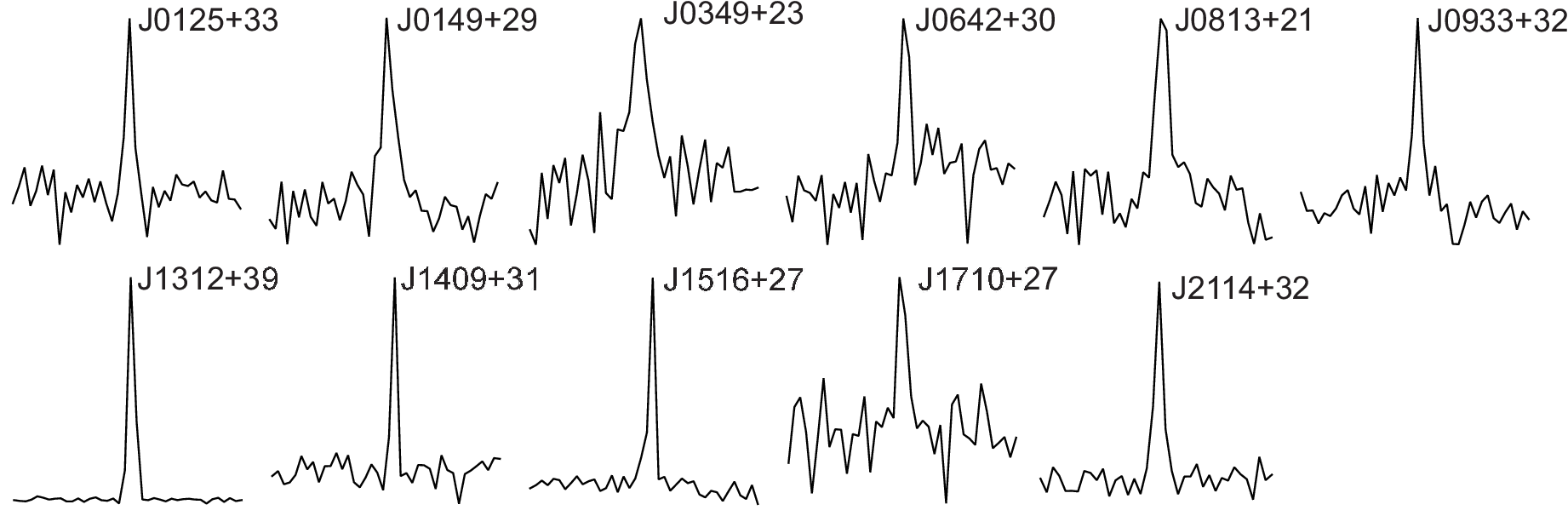}
    \caption{RRAT pulse profiles are shown. The name of the transient is displayed in the upper right corner of the profile. Other characteristics of the presented pulses: P, DM, $W_{e}$ and $S_{p}$ in the Table~\ref{tab:tab1}. For each profile, the recording duration is 0.5 s.}
   \label{fig:fig2}
\end{figure*}

\section{Results}

Upon visual inspection, it turned out that for 69 candidates out of 116 (59.4\%), it was not possible to find pulses with a S/N > 6. Their nature remained unknown. 21 candidates (18.1\%) show a signal with a dispersion delay, but his signal is observed in less than half of the frequency channels, and the line of dispersion delay does not obey the quadratic dependence of the pulse arrival time on frequency. Apparently, these signals are associated with interference of an unknown nature. 15 sources (12.9\%) were identified with bright pulsars observed in the far
side lobes of the antenna. For example, candidates J1505+4207, J1506+35, J1508+28, J1511+30, J1511+34 with detection of several thousand pulses with the same DM were found. All these candidates are identified with the bright pulsar J1509+5531 (B1508+55) located outside the studied site ($+21^o < \delta < +42^o$) and having a direct relationship close to the candidates found. Taking into account that B1508+55 has previously been detected dozens of times in other side lobes, we can be sure that it has been rediscovered in the far side lobes of the LPA.

Eleven candidates (9.5\%) turned out to be new RRAT. Examples of dynamic spectra of two RRAT is shown in Fig.\ref{fig:fig1}. J0813+21 transient having two visible pulses located at a distance 42 pixels (525 ms). On the dynamic spectrum J1312+39 shows a very bright pulse, which was the brightest for all 11 transients found. On the website (https://bsa-analytics.prao.ru/en/) in the search for pulsars and transients at the LPA LPI, the dynamic spectra of all found transients are given. Fig.\ref{fig:fig2} shows the pulse profiles of the found RRAT. 

\begin{table*}
	\centering
	\caption{Some characteristics of the found RRAT}
	\label{tab:example_table}
	\begin{tabular}{lcccccccc}
	\hline
	Name & $\alpha_{2000}^{(h,m,s)}$ &	$\delta_{2000}^{(o,')}$ & P (s) & $DM$(pc/cm$^3$) &	$W_{e}$ (ms) &	$S_{p}$ (Jy) &	$S/N$ &	$N$ \\
	\hline
J0125+33 & 01 25 50	& 33 10 & -	& 21.0	& 15 & 3.3 & 9.6 & 3 \\
J0149+29{$^\ast$} & 01 49 00 & 29 12 & 2.654 & 36.5	& 30 & 3.1 & 8.5 & 3 \\
J0349+23 & 03 49 30	& 23 42 & -	& 58.5	& 40 & 2.9 & 8.0 & 4 \\
J0642+30 & 06 42 50	& 30 37 & 1.4114 & 39.5	& 25 & 2.7 & 8.1 & 1 \\
J0813+21{$^\ast$} & 08 13 30 & 21 54 & 0.531 & 51.5	& 35 & 3.1 & 6.7 & 4 \\
J0933+32{$^\ast$} & 09 33 10 & 32 54 & 0.9616 & 18.0 & 25 & 8.5 & 10.2 & 1 \\
J1312+39 & 13 12 40	& 39 56 & -	& 12.5 & 15 & 165 & 210 & 3 \\
J1409+31 & 14 09 50	& 31 35 & -	& 8.0 & 20 & 4.2 & 12.1 & 4 \\
J1516+27 & 15 16 20	& 27 59 & 1.125	& 14.5 & 20 & 12.8 & 29.4 & 24 \\
J1710+27 & 17 10 20	& 27 01 & -	& 17.5 & 25 & 3.1 & 8.4 & 1 \\
J2114+32 & 21 14 10	& 32 18 & 0.597	& 25.5 & 20 & 4.7 & 10.7 & 2 \\
	\hline
 \label{tab:tab1}
\end{tabular}
\end{table*}

Table~\ref{tab:tab1} shows the measured parameters of the found RRAT. Columns 1-3 give the name and coordinates of the transients for 2000, columns 4-9 show the period (if defined), the dispersion measure, the half-width of the average profile ($W_{e}$), the observed peak flux density ($S_{p}$) for the brightest pulse, the pulse width, the number (N) pulses with S/N>7. The accuracy of the coordinates in right ascension was determined as half the size of the radiation pattern ({$\pm$}1.7m), the accuracy of the coordinates in declination was half the distance between adjacent beams ({$\pm15'$}). The period was defined as the largest total time interval when observing several pulses per day or from the power spectra. The current period may be an integer number of times less than the value indicated in the table. The typical accuracy of determining the period is 0.005 s. To determine the DM, the dependence of the S/N pulse on the iterated DM was built. The accuracy of the definition {$\pm$}2.0 pc/cm$^3$. The half-width of the profile was determined at half the height of the pulse. The peak flux density was determined from the background temperature in the direction of the transient and the observed S/N. When obtaining the estimate, the zenith distance of the source and the correction for the signal response in the beam were taken into account. For almost all transients, the $S_{p}$ estimate is the lower estimate of the pulse flux density, since we do not know the exact coordinate of the transient, and it is not possible to make corrections that take into account the transient's location outside the center of the radiation pattern. Exact corrections were made for J0813+21 and J0933+32, since their coordinates are known from ATNF, as well as for J1312+39, the pulses from which are observed in two adjacent beams at almost identical S/N. The actual peak pulse flux densities for the remaining transients can be up to 1.5–2 times higher than the values given Table~\ref{tab:tab1}.

The asterisk in Table~\ref{tab:tab1} marks transients J0149+29, J0813+21, and J0933+32, which previously had regular radiation. RRAT J0149+29 (DM = 34.5 pc/cm$^3$; P = 2.654 s) is discussed in our work on the search for pulsars, which is under survey (Tyulbashev et al., sent to MNRAS). In the table. 1 we give a period description from this work. Pulsar J0813+22 (DM = 52.29 pc/cm$^3$; P = 0.5314 s) was found in the LOTAAS survey conducted at 135 MHz on LOFAR (\citeauthor{Sanidas2019}, \citeyear{Sanidas2019}). The authors cite an average profile with an observed S/N of 5-6, but do not note the detection of individual pulses, although they searched for dispersed pulses. The period shown in the table is defined independently as the distance between the recorded pulses. Pulsar J0935+33 (DM = 18.35 pc/cm$^3$; P = 0.9615 s) was detected on LPA3 earlier (\citeauthor{Tyulbashev2017}, \citeyear{Tyulbashev2017}), but its individual pulses were not recorded. The period shown in the table is determined by the distance between individual pulses. Note also that, according to formal criteria, the source J0813+21 must be excluded from Table 1, since its brightest pulse has a S/N < 7. However, the coincidence of coordinates, DM and P with the known pulsar allows it to be left in the table.

\begin{figure*}
	\centering
	\includegraphics[width=\textwidth]{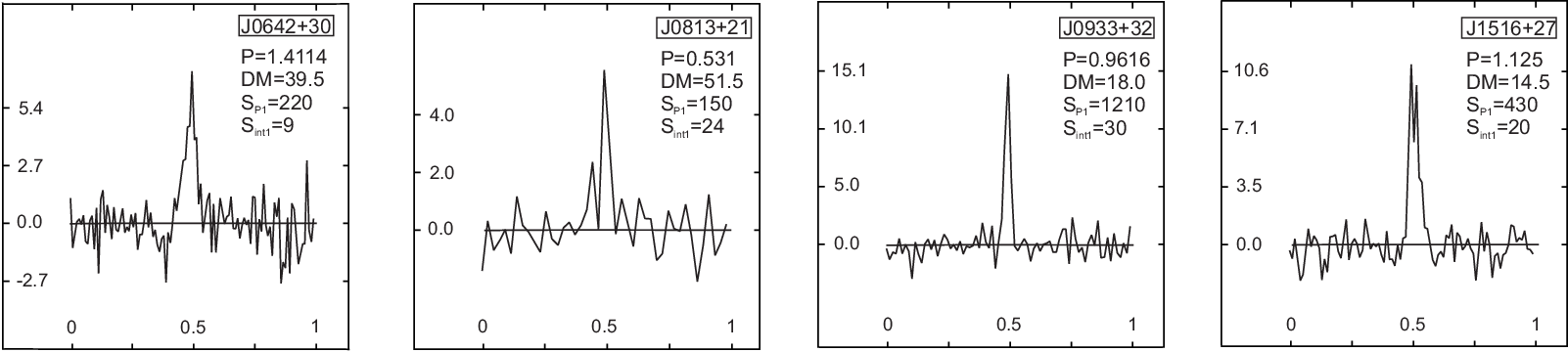} 
	\caption{Average profiles accumulated in one session over 180 s. According to the vertical axis of the profile, according to the horizontal - the phase. In the upper right corner of the average profiles is the name of the transient, P (in seconds), DM (pc/cm$^3$), $S_{p1}$ and $S_{int1}$ (in mJy).}
    \label{fig:fig3}
\end{figure*}

A search for regular (periodic) radiation was carried out for the found transients. The average profiles were obtained for four RRAT (see Fig.\ref{fig:fig3}).

\section{DISCUSSION OF THE RESULTS AND THE CONCLUSION}

1. For transients J0642+30 and J1516+27, the average profiles were constructed according to the data in one observation session. At the same time, their periodic radiation was not detected in the search for pulsars using summed power spectra, which was carried out over an interval of two to seven years (Tyul'bashev et al., sent to MNRAS). The sensitivity in the Pushchino Multibeams Pulsar Search (PUMPS) is 35-40 times higher than the sensitivity in one observation session, and reaches 0.2 mJy (\citeauthor{Tyulbashev2022}, \citeyear{Tyulbashev2022}). In the presented average profiles, the peak and integral flux densities of J0642+30 and J1516+27 are, respectively, $S_{p1}$ = 220 and 430 mJy, $S_{int}$ = 9 and 20 mJy. Since the average profiles were accumulated in three-minute sessions, and at the same time J0642+30 and J1516+27 were not detected when averaging spectra over 3000 days for one session per day, then these RRATs should have very strong variability.\\
2. The transients J0642+30, J0813+21, J0933+32, J1516+27 have the usual pulsar radiation. The ratio of the peak flux density of the brightest pulses and the peak flux density in the average profile $S_{p}/S_{p1}$ = 12, 21, 7, 30. We believe that these transients are ordinary pulsars with a long tail of pulse energy distribution, or pulsars with giant pulses (J0813+21, J1516+27). Observations with sensitivity several times higher than the sensitivity of LPA3 can show this unequivocally. \\
3. The number of transient pulses found with S/N>7 varies from 1 to 24. With the exception of J1516+27, it does not exceed four pulses found during a time equivalent to an observation session lasting six days. Thus, to confirm these transients, it takes from one and a half to six days of observations for each source on antennas with sensitivity
equivalent to the sensitivity of LPA3.\\
4. The most interesting of the transients found is RRAT J1312+39. It has only three pulses detected at S/N > 7, but all the pulses are bright. Their S/N = 210, 60, 42, and peak flux densities, respectively, $S_{p}$ = 165, 46, 33 Jy. In the signal level range 6 < S/N < 7 not a single pulse was detected. With the accumulation of power spectra over 3000 sessions, the periodic signal is not detected.  According to the work of \citeauthor{Tyulbashev2022} (\citeyear{Tyulbashev2022}), the upper estimate of the integral flux density of this transient should be < 0.5 mJy if its period is 0.5 < P < 3 s. A long tail of the energy distribution of pulses for this transient is unlikely, since in this case, the more pulses should be observed, the smaller their S/N. If J1312+39 is a pulsar with nulling, then the proportion of nulling can be estimated by assuming a pulsar period of 1 s and 10 s. In this case, the share of nulling will be from 99.999 to 99.994\%. J1312+39 has the brightest pulse of all the RRATs found on the northern hemisphere. Previously, the brightest pulse was observed at J0139+33 (40 Jy; \citeauthor{Tyulbashev2018b} (\citeyear{Tyulbashev2018b}). The $S_{p}$ of the pulse of J1312+39 is comparable to the estimates of the $S_{p}$ pulses of the brightest second pulsars observed earlier in the search for RRAT on LPA3 (\citeauthor{Tyulbashev2018a}, \citeyear{Tyulbashev2018a}): J1136+1551 (B1113+16; 408 Jy), J1239+2453 (B1237+25; 215 Jy), J0953+0755 (B0950+08; 174 Jy) and J0837+0610 (B0834+06; 107 Jy). With the exception of  B0834+06, the remaining sources are on the lists of pulsars with giant pulses (\citeauthor{Kazantsev2018}, \citeyear{Kazantsev2018}).

The main result of the work during the verification of the previously detected weak pulses in the archival data accumulated over eight years is the discovery of 11 new RRATs. Periods are estimated for six transients. The total number of found RRATs in the PUMPS survey (\citeauthor{Tyulbashev2022}, \citeyear{Tyulbashev2022}) reached 59.

\section*{Acknowledgements}
The study was carried out at the expense of a grant Russian Science Foundation 22-12-00236\footnote{https://rscf.ru/project/22-12-00236/}. The authors are grateful to L.B. Potapova for help with the preparation of some figures.

\end{document}